# A coordinated control to improve performance for a building cluster with energy storage, electric vehicles, and energy sharing considered

Pei Huang[1*], Marco Lovati[1], Xingxing Zhang[1], Chris Bales[1]

[1] Department of Energy and Built Environment, Dalarna University, Falun, 79188, Sweden
[*]Corresponding author. Tel.: +46 (0) 23-77 87 89; E-mail: phn@du.se

**Abstract:** Distributed renewable energy systems are now widely installed in many buildings, transforming the buildings into 'electricity prosumers'. Existing studies have developed some advanced building side controls that enable renewable energy sharing and that aim to optimize building-cluster-level performance via regulating the energy storage charging/ discharging. However, the flexible demand shifting ability of electric vehicles is not considered in these building side controls. For instance, the electric vehicle charging will usually start once they are plugged into charging stations. But, in such charging period the renewable generation may be insufficient to cover the EV charging load, leading to grid electricity imports. Consequently, the building-cluster-level performance is not optimized. Therefore, this study proposes a coordinated control of building prosumers for improving the cluster-level performance, by making use of energy sharing and storage capability of electricity batteries in both buildings and EVs. An EV charging/discharging model is first developed. Then, based on the predicted future 24h electricity demand and renewable generation data, the coordinated control first considers the whole building cluster as one 'integrated' building and optimizes its operation as well as the EV charging/discharging using genetic algorithm. Next, the operation of individual buildings in the future 24h is coordinated using nonlinear programming. For validation, the developed control has been tested on a real building cluster in Ludvika, Sweden. The study results show that the developed control can increase the cluster-level daily renewable self-consumption rate by 19% and meanwhile reduce the daily electricity bills by 36% compared with the conventional controls.

# 1. Introduction

Buildings represent large energy end-users worldwide [1]. In the E.U. and U.S, buildings currently consume over 40% of total primary energy usage [2]. Renewable energy, which has much less carbon emissions and relatively lower costs compared with the conventional fossil fuel-based energy, offers a promising solution to meeting the large energy needs in the building sectors [3]. In this regard, distributed energy systems, such as PV panels, have gained popularity and are now widely installed in buildings [4]. For instance, the Swedish Energy Agency has set a target for 100% renewable electricity production by 2040, to which building integrated PV systems are planned to contribute 5-10% in electricity generation [5]. The integration of distributed energy systems has promoted the transformation of buildings' role from energy consumers to energy prosumers, i.e. energy consumers who produce energy for their own consumption using distributed energy technologies [6]. A popular type of energy prosumer is the zero energy buildings (ZEBs) [7], which produce the same amount of energy as they consume. The transformation of buildings' role into energy prosumers also provides opportunities for collaborations among buildings to improve the overall cluster-level performances [8]. When multiple building prosumers are involved in a building cluster, they can share their excessive renewables with others with insufficient generations [9]. Such energy sharing can help improve the building-cluster-level renewable self-consumption rates and thus reduce the grid power usage (due to an increased share of renewable energy utilization). A study conducted by Luthander et al. [10] shows that that even a simple energy sharing (i.e. aggregate electricity demand and supply) among 21 houses in Sweden can easily improve the PV power self-consumption by over 15%. When there is shared energy storage, the improvement in PV power self-consumption can reach 29%.

To achieve energy sharing among buildings, existing studies have developed a number of advanced controls. For example, Odonkor et al. proposed a control method of ZEBs using genetic algorithm and Pareto decision making based on an adaptive bi-level decision model (with a facilitator agent at cluster level and local systems at single NZEB level) [11]. In such bi-level decision model, the individual building's systems (i.e. individual PV system and battery) are in the first level, and the centralized cooling system as well as an ice storage system are in the second level. Fan et al. proposed a collaborative demand response control of zero energy buildings for enhancing the building-cluster-level performances. In their method, the control of each building was conducted in sequence, and the optimization of one building's operation was based on the previously optimized buildings' operation [9]. In each optimization, the daily hourly charging/discharging rates of the battery are set as variables to be optimized, and the economic cost and grid friendliness are set as the objective function. Prasad and Dusparic developed a Deep Reinforcement Learning based method for ZEB community [12]. The ZEB community is modelled as a multiagent environment, where each agent represents a building. Every agent learns the optimal behaviour independently and is entirely responsible for making energy transactions on behalf of that building. The abovementioned controls optimize the building cluster performance in a bottom-up way, and they merely perform very limited collaborations among buildings.

With the purpose of maximizing the energy sharing within a building cluster, researchers have developed controls that directly use the building-cluster-level performances as the optimization targets. For instance, Gao et al. [13] developed a genetic algorithm based coordinated demand response control in which all the storage systems' charging rates were optimized simultaneously. Similarly, considering the problems caused by the independent microgrids operation, Zhang et al. [14] proposed a coordinated control in which a cluster-level controller (i.e. an aggregator) was utilized to simultaneously manage local energy transactions among microgrids and energy exchanges with the grid. Both these two controls are easy to implement and effective in improving building-cluster-level performance. But with the increase of the number of buildings, such a straightforward coordination will face too many parameters to be optimized, causing excessive computation load. Such excessive computation loads make the straightforward method unfeasible to be applied in large NZEB clusters. To address the large computation, Huang et al. developed a top-down control for a cluster of building prosumers equipped with electrical energy storage system [15]. In their study, the optimal performances that can be achieved are first searched by advanced searching algorithm. Then the optimal performances at the top-level are divided into separate goals for each individual building at the bottom-level. Compared with the individual controls, their method can increase the daily load coverage by renewable energy by as much as 45%, reduce the daily peak energy exchanges with the power grid by as much as 80%, and meanwhile significantly reduce the daily operational costs. Similarly, in [16] a three-step demand response control algorithm is developed considering the dynamic pricing. Such control can flatten the electricity demand profiles via properly coordinating single buildings and thus maximize the benefits of both buildings and the power grid. Taking into account of the demand prediction uncertainty, in [17] a robust collaborative control is developed. Such control identifies the optimal operation strategy under the predicted uncertain ranges of demand, and thus it can maximize the renewable energy sharing robustly.

These existing controls can effectively improve the performances at building cluster level. However, electric vehicles (EV), which also play an important role in the building cluster scale energy systems, are usually considered as non-scheduled electrical loads (such as lighting) and their flexible demand shifting ability is rarely used [18, 19]. As a result, the flexible demand shifting ability of EVs are rarely considered together with the building control, leading to limited performance improvements at building cluster level [20] [21]. For instance, in practice the EV charging will start once they are plugged into charging stations. However, in such charging period the renewable generation may be insufficient to cover the EV charging load, leading to grid electricity imports. On the other hand, when there is surplus renewable generation, the EVs cannot be used as electricity storage if they have already been fully charged, leading to the surplus renewable energy exports. As a result, the overall building-cluster-level performance is not fully optimized.

The EV deployment is continuously increased and many governments have established policy or goals to promote the EV deployment. For instance, the French government set a target of 2 million EVs in 2020 [22]. The Swedish government has set a goal that the vehicle fleets should be 100% independent of fossil fuel by 2030 (a large percentage should be achieved by EV deployment) [23]. The U.S. Federal government has enacted policies and legislations to promote the U.S. market for EVs, such as improvements of tax credits in current law, and competitive programs to encourage communities to invest in infrastructure supporting these vehicles [24]. The number of EVs on the road is projected to reach 18.7 million in 2030, up from slightly more than 1 million at the end of 2018 [25]. In the future, due to the large penetration, EVs will have large impacts on the grid power demands. Thus, it is necessary to make use of their potentials in demand regulation in the power grid and district energy systems.

By properly scheduling the EV charging loads, the batteries in EVs can be used as flexible energy storage to help regulate the electricity demands in the power grid. Existing studies have also developed some advanced controls for EVs at both individual level and aggregated level. At individual level, Islam et al. proposed a coordinated EV charging control based on a correlated probabilistic model of EV charging loads considering the stochastic charging behaviour [26]. The charging control optimizes the power factors of PV and battery energy storage system to enhance the quality of service by minimizing the probability of voltage and current noncompliance. The application of the developed control on a three-phase IEEE 37-bus unbalanced distribution system using the real data of vehicles and solar PV shows it is effective in providing more quality of service. At aggregated level, Geth et al. developed a coordinated charging control for a number of EVs [27]. In their control, a vehicle owner first indicates the point in time when the batteries should be fully charged. Then, the aggregator collects this information and calculates when each EV can start charging, based on two rules: *(i)* charging is most economically when the total demand (including the residential, industrial and EV consumption) is low, and *(ii)* the EVs can be charged during working hour in the working places. Case study shows that the coordinated charging can effectively decrease the peak load, as the coordination makes the charging load profile much flatter. Similarly, Usman et al. proposed an automated coordinated control of EV fleets, which can plan the charging strategy at the cheaper moments while keeping the vehicle charged enough to complete its scheduled trips [28]. Their control uses a grid agent to grant tokens to the EVs in idle state based on the grid electricity prices. By shifting charging loads to low electricity price period (usually with low aggregated electricity demands in the power grid), this control can effectively increase the match between the available power and the consumed power. In [29], three different smart EV charging control methods were proposed for increasing PV power usage in a microgrid: real-time charging based on PV power sufficiency; real-time charging based on PV power sufficiency and with vehicle-to-grid enabled (i.e. vehicle can discharge power back to grid); and linear programming based optimization (i.e. optimize all EVs' charging based on PV power production on a daily basis). Their study shows that coordinated EV controls can increase the PV power self-consumption by 13~38% and reduce the peak electricity demand by 27~67%. In this study, the optimization at the building side (e.g. battery charging/discharging and energy sharing) was not considered. Fachrizal and Munkhammar [30] developed a centralized smart EV charging scheme for a residential building cluster, which optimizes the charging rates of all the EVs simultaneously considering the interaction of individual EVs and PV power production. This study innovatively considered EV smart charging in improving building-cluster-level performance. However, their study considers all the buildings in the cluster as one 'aggregated' building for simplicity, in which the operation of individual buildings is neglected. Such a simplification may not be practical. The abovementioned studies can effectively improve the economic performances of EV or EV fleets. However, these studies typically consider EVs as a separate role in the urban energy system and thus neglect their integration with the building controls. In the future scenario with increased number of building prosumers and EV penetration, EV control not integrated with building prosumers controls (i.e. buildings' energy sharing control) will limit the overall performance improvement potentials.

To sum up, existing studies have developed some advanced building side controls that enable renewable energy sharing and that aim to optimize building-cluster-level performance via regulating the energy storage's charging/discharging. However, the flexible demand shifting capability of EVs, which has been proven effective in enhancing building cluster-level performance (e.g. increase PV power self-consumption by over 10% [30], reduce peak demand by 37% [31]), is not considered in the cluster-level controls. Therefore, this study proposes a coordinated control of building cluster with both energy sharing and the EV charging considered, with the purpose of improving the cluster-level performance by taking advantage of energy sharing and storage capability of electricity batteries in both buildings and EVs. An EV charging/discharging model is first developed, and then a coordinated control is developed for building cluster with the energy storage, EVs and energy sharing considered. Based on the predicted future 24h electricity demand and renewable generation data, the coordinated control first considers the whole building cluster as one 'integrated' building and optimizes its operation as well as the EV charging/discharging using genetic algorithm. Then, the operation of individual buildings in the future 24h is coordinated using nonlinear programming. For validation purpose, the developed control has been tested using the energy demand and supply data on a real building cluster in Ludvika, Sweden. The major contributions of this study to the subject have also been summarized and added to the introduction.

- A coordinated control method for building clusters has been developed for optimizing building-cluster-level performances.
- The developed control can regulate individual building's battery charging/discharging to maximize the renewable energy sharing and at the same time coordinate all the individual EVs' charging load in order to increase the building cluster renewable energy usage.
- The performances of the developed control have been compared with a conventional individual control (without energy sharing and smart EV charging) and a partial collaborative control (with energy sharing but without smart EV charging).
- The performance improvements have been analyzed in aspects of renewable energy self-consumption and economic saving.

The structure of the paper is as followings. Section 2 describes the overall coordinated control for the building cluster. Section 3 presents the detailed building model and energy system models. In section 4, the developed coordinated control is applied on a case building cluster. and its performance are compared with two existing scenarios. The brief conclusions are given in section 5.

## 2. Coordinated control to improve energy performance for a building cluster

This section first introduces the energy sharing concept. Then, the detailed coordinated control, which takes account of the energy storage, EVs charging and renewable energy sharing among buildings, is introduced.

### 2.1 Energy sharing

Energy sharing is an effective way to improve the overall performances at the building cluster level. In this study, the energy sharing is implemented by installing an energy sharing microgrid among the buildings [19] [32], as depicted by Fig. 1. The renewable energy from Building A can be used to supply the electricity demands charge the EVs in Building B or C, or even be stored in the battery of Building B or C. Such renewable energy sharing can help increase the renewable self-utilization rates of the building cluster, and thus help improve both the economic and energy performances [15].

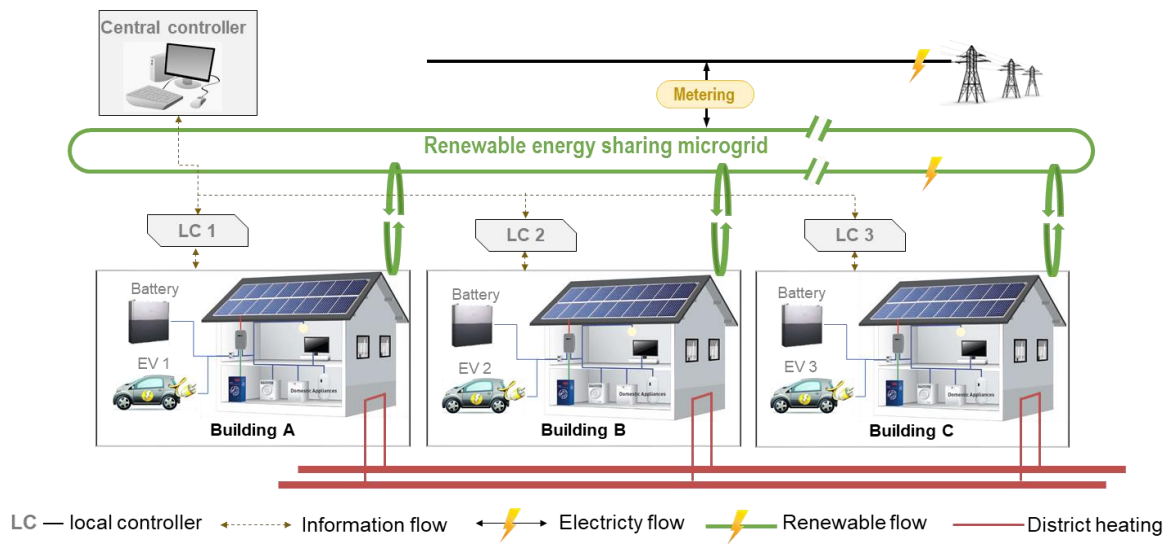

Figure 1 Schematics of electricity energy sharing among buildings in a cluster

### 2.2 A coordinated control to improve energy performance for a cluster of building energy prosumers with energy storage, EVs, and energy sharing considered

This section introduces the developed coordinated control. Fig. 2 presents the flowchart of the developed method. The aim of the coordinated control is to coordinate the operation of energy storage (installed in each single building) and the EVs, to achieve the optimal cluster-level performances. The coordinated control consists of four steps. In Step 1, all the buildings in the building group are considered as a 'representative' building, and the electrical demand, renewable energy generation and load shifting capacity of the 'representative' building are predicted, i.e. its electrical demand/renewable generation/demand shifting capacity equals the aggregated demand/ generation/capacity of all buildings inside the cluster. In Step 2, the operation of the 'representative' building and the EV charging rates are optimized using genetic algorithm (GA). The performance of the 'representative' building, obtained by simultaneous optimization of the building and EV operation, is considered to be the best performances that the building group can achieve [33]. In Step 3, the operation of each single building inside the building group is coordinated using non-linear programming (*NLP*) based on the 'representative' building's operation obtained from Step 2. In Step 4, the performances of the proposed coordinated control are compared with two existing controls, including a conventional individual control (Scenario 1 [33]), which does not enable renewable sharing and charge the EVs immediately after being parked, and an existing coordinated control (Scenario 2 [13]), which enables full renewable energy sharing but also charges the EVs immediately after being parked. The details of each step are introduced below.

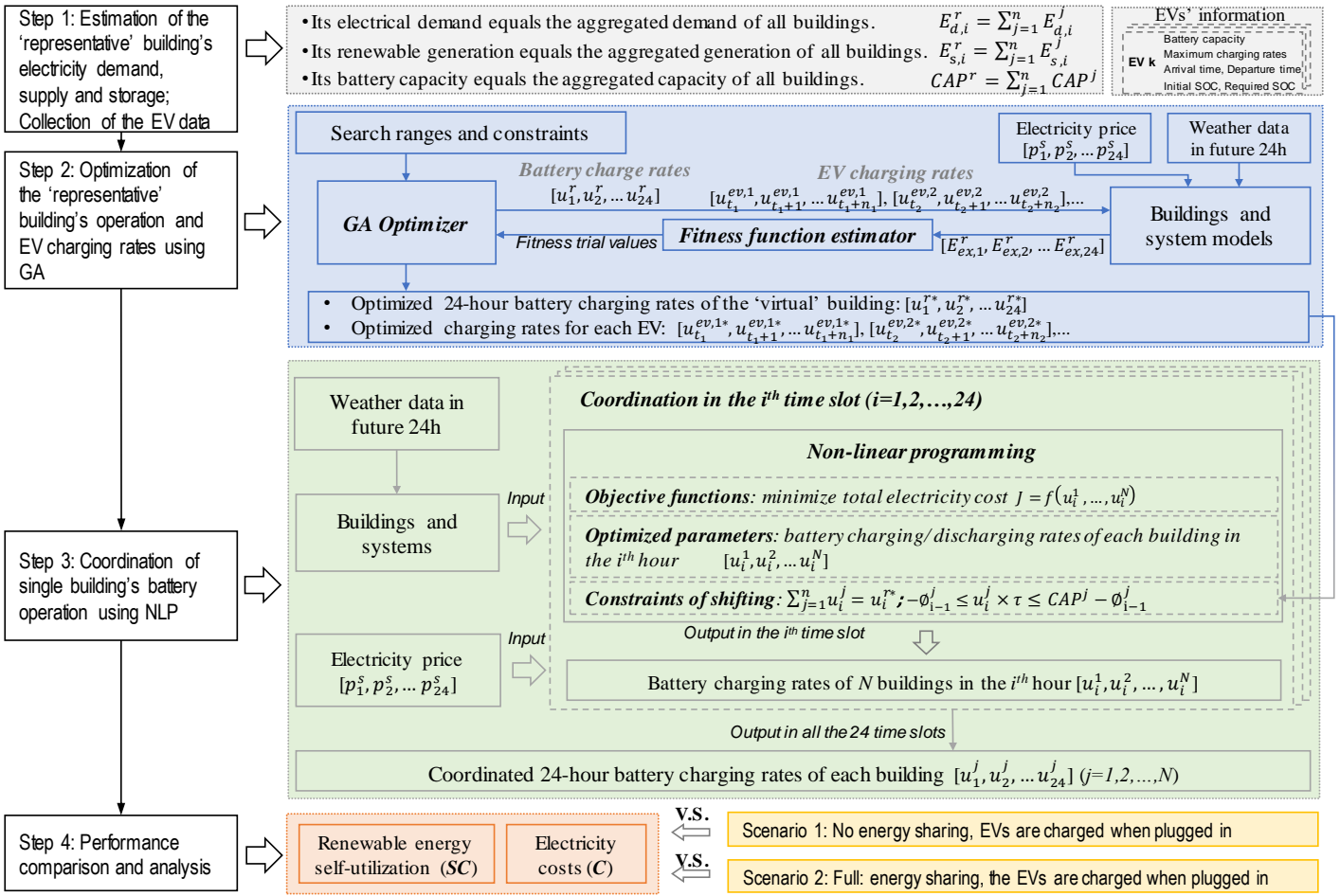

Figure 2 Flowchart of the coordinated control to improve energy performance for a building cluster with energy storage, EVs, and energy sharing

*Step 1: Estimation of the 'representative' building's demand and storage*

In this step, all the buildings inside the cluster are considered as a 'virtual' building. Its hourly electricity demand ($E_{d,i}^r$ (kW·h)) equals the aggregated hourly electricity demand of each single building ($E_{d,i}^j$ (kW·h)) ($i$ indicates time with a unit of hour), its hourly renewable generation ($E_{s,i}^r$ (kW·h)) equals the aggregated hourly renewable generation of each single building ($E_{s,i}^j$ (kW·h)) and its load shifting capacity ($CAP^r$ (kW·h), i.e. battery capacity) is the aggregated load shifting capacity of each single building ($CAP^j$ (kW·h)).

$$E_{d,i}^r = \sum_{j=1}^n E_{d,i}^j \quad (1)$$

$$E_{s,i}^r = \sum_{j=1}^n E_{s,i}^j \quad (2)$$

$$CAP^r = \sum_{j=1}^n CAP^j \quad (3)$$

*Step 2: Optimization of the 'representative' building's operation using GA*

The electricity demand and renewable generation of each individual building is calculated using the models presented in Section 3. The GA algorithm searches the optimal charging/discharging rates of both the battery (see Section 3.2 for the detailed models) and EVs that can minimize the electricity costs of the 'representative' building. For example, the EVs can be scheduled to be charged in periods with sufficient renewable generations while not charged in periods with insufficient generations. In the GA simulation, the inputs mainly include the battery charging/discharging rates (to be optimized), the EV charging rates (to be optimized), the EV parking periods, the future 24-hour weather data, building parameters, and battery parameters. The EVs are different from the electrical battery, since they are not constantly connected into the buildings. This study uses four parameters to characterize an EV (e.g. the $k^{th}$ EV): arrival time to the charging port ($t_k$), parking periods in the charging port ($n_k$), initial state of charge ($SOC_{0,k}$), and the required state of charge when the car departs from the charging port ($SOC_{1,k}$). These parameters are considered known and will be used as inputs in the optimization.

In each generation of GA, trials of 24-hour thermal storage hourly charging/discharging rates (i.e., $[u_1^v, u_2^v, ... u_{24}^v]$ kW) and charging rates of each EV (i.e. $[u_{t_k}^{ev,k}, u_{t_k+1}^{ev,k}, ... u_{t_k+n_k}^{ev,k}]$ kW) are generated by the GA optimizer. The representative building's hourly power demand ($E_{d,i}^r$ kW) and hourly renewable power generation ($E_{r,i}^r$ kW) in the future 24 hours is then predicted using the building and system models (see models given in Section 3). The charging/discharging rates of the electrical battery should meet the following two constraints: *(1)* The battery charging amount could not exceed the remaining battery storage capacity. *(2)* The battery discharging amount could not exceed the stored electricity in the battery. These two constraints are expressed by Eqn. (4) [34, 35],

$$0 \leq \emptyset_0^r + (u_1^r + u_2^r + \cdots + u_i^r) \times \tau \leq CAP^r \quad \text{where } i=1,2,...,24 \quad (4)$$

where $\emptyset_0^v$ ($kW·h$) is the amount of thermal energy initially stored in the tank, $\tau$ is the duration of battery charging/discharging (i.e., 1 hour in this study).

Similarly, the charging rates of the $k^{th}$ EV should meet these two constraints, as expressed by Eqn. (5). $SOC_{0,k}$ is the initial state of charge when the $k^{th}$ EV arrives at the charging port. $CAP_k^{ev}$ ($kW·h$) is the capacity of the $k^{th}$ EV battery. $t_k$ is the arrival time of the $k^{th}$ EV at the charging port, and $n_k$ is the parking duration.

$$0 \leq SOC_{0,k} \times CAP_k^{ev} + (u_{t_k}^{ev,k} + u_{t_k+1}^{ev,k} + u_{t_k+i}^{ev,k}) \times \tau \leq CAP_k^{ev} \quad \text{where } i=1,2,\ldots, n_k \quad (5)$$

In addition, the EV battery should be charged to a user-specified level ($SOC_{1,k}$) before they depart the charging port. This constraint is expressed by Eqn. (6). When $SOC_{1,k}$ equals 1, it represents the EV users require the EV battery to be fully charged before they depart the charging port.

$$SOC_{0,k} \times CAP_k^{ev} + (u_{t_k}^{ev,k} + u_{t_k+1}^{ev,k} + \cdots + u_{t_k+n_k}^{ev,k}) \times \tau \geq SOC_{1,k} \times CAP_k^{ev} \quad (6)$$

This study considers the strategy to minimize daily electricity cost of the building group. Following this control goal, a fitness function is determined, as expressed by Eqn. (7) [36].

$$J_{grid} = \min(Cost) \quad (7)$$

$$Cost = \sum_{i=1}^{24} E_{ex,i}^r \times \tau \times \chi_i, \begin{cases} \chi_i = \chi_{buy}, \text{if } E_{ex,i}^r > 0 \\ \chi_i = \chi_{sell}, \text{if } E_{ex,i}^r \leq 0 \end{cases} \quad (8)$$

where $\chi_i$ ($kr/(kW·h)$) is the electricity price in the $i^{th}$ time slot. $\chi_{buy}$ ($kr/(kW·h)$) is the price of purchasing electricity from the power grid, and $\chi_{sell}$ ($kr/(kW·h)$) is the feed-in-tariff.

The outputs of the GA search are the 'representative' building's battery charging/discharging rates ($[u_1^{r*}, u_2^{r*}, \ldots u_3^{r*}]$ $kW$) in the next 24 hours and the charging rates of each individual EV ($[u_{t_1}^{ev,1*}, u_{t_k+1}^{ev,k*}, \ldots, u_{t_k+n_k}^{ev,k*}]$, $[u_{t_2}^{ev,2*}, u_{t_2+1}^{ev,2*}, \ldots u_{t_2+n_2}^{ev,2*}], \ldots kW$). The optimized battery charging/discharging rates of the 'representative' building are used in Step 3.

*Step 3: Coordination of single building's operation using NLP*

In this step, the single building's battery charging/discharging rates (i.e. $u_i^j$ is the $j^{th}$ building in the $i^{th}$ hour) are coordinated using NLP based on the 'representative' building's operation [37]. The NLP is conducted in each hour and will be repeated 24 times for obtaining the building's daily operation. The fitness function of the NLP is expressed by Eqns. (6) and (7), which aims at minimizing the electricity costs of the building group.

$$J_{NLP} = \min(Cost_{all,i}) \quad (9)$$

$$Cost_{all,i} = \sum_{j=1}^{n} (E_{d,i}^j \times \chi_i)^2 \quad (10)$$

In order to reduce the uneven allocation of the battery charging/discharging rates (otherwise only a few buildings take benefits from the demand response), the square of each building's operational cost is used in the fitness function. $E_{d,i}^{j}$ (kW·h) is the energy demand of the $j^{th}$ building in the $i^{th}$ hour after applying the $u_i^j$ (kW) amount of battery charging/discharging, which is calculated by the models presented in Section 3. $\chi_i$ (HKD/(kW·h)) is the electricity price in the $i^{th}$ hour.

In the $i^{th}$ hour, the optimized parameters in the NLP are the hourly battery charging/discharging rates of all the buildings inside the building group, i.e., $[u_i^1, u_i^2, \ldots, u_i^N]$ (kW), where $N$ indicates the number of buildings in the building group. The battery charging/discharging rates in each hour should follow the constraints below.

(i) The sum of battery charging/discharging rates of each building ($u_i^j$ (kW)) should equal the battery charging/discharging of the 'representative' building ($u_i^{r*}$ (kW)) (obtained from Step 2).

$$\sum_{j=1}^{N} u_i^j = u_i^{r*} \qquad (11)$$

(ii) For each single building, the electricity charging amount must be smaller than the remaining storage capacity of the battery, and the electricity discharging amount must be smaller than the amount of electricity stored in the battery. There are $2N$ inequality constraints for $N$ buildings.

$$-\emptyset_{i-1}^{j} \leq u_i^j \times \tau \leq CAP^j - \emptyset_{i-1}^{j} \quad \text{(j=1,2,...N, respectively)} \qquad (12)$$

where $\tau$ is the charging duration (i.e., 1 hour), $CAP^j$ (kWh) is the battery capacity of the $j^{th}$ building, $\emptyset_{i-1}^{j}$ (kW·h) is the electricity energy stored in the $j^{th}$ building's battery. $\emptyset_{i-1}^{j}$ (kW·h) is calculated by Eqn. (10).

$$\emptyset_{i-1}^{j} = (u_1^j + u_2^j + \cdots + u_{i-1}^j) \times \tau \qquad (13)$$

In total, for a building group with $N$ buildings, there are $N$ unknown parameters to be solved, there is 1 equality constraints, and there are $2N$ inequality constraints.

*Step 4: Performance comparison and analysis*

After obtaining the optimized operation of each single building, the performances of the proposed coordinated control are compared with two existing controls in aspects of renewable energy self-consumption improvements and economic cost savings. The two existing controls include a conventional individual control (Scenario *1* [33] [13]), which does not enable renewable sharing and charge the EVs immediately after connecting them, and an existing coordinated control (Scenario *2* [15]), which enables full renewable energy sharing in the building cluster but charges the EVs immediately after connecting them. In both the two comparative studies, the EVs demand are first computed. Such load is added to the building electricity demand, which will then be used as inputs for battery charging/discharging controls. In Scenario 1 (i.e. an existing individual control) [33], GA was used for searching the optimal battery charging/discharging rates in each building, which is similar to the control optimization of the 'representative' building (see Step 2 in Fig. 2 without EV related variables). After obtaining the individual buildings' optimal operation, their electrical demands were aggregated for evaluating the building-cluster-level performances. In Scenario 2 (i.e. an existing coordinated control) [13], the battery charging/discharging rates of all the three buildings are optimized simultaneously using GA, and the minimization of the building-cluster-level performance was used as the fitness function.

Table 1 Configuration of the three scenarios

| Scenario | EV control? | Energy sharing? |
|---|---|---|
| 1 | Charged immediately when plugged in | No |
| 2 | Charged immediately when plugged in | Full sharing |
| 3 (Developed control) | Charged at any time when parked | Full sharing |

## 3. Buildings and system modelling

This section introduces the building information and system modelling. Each building is installed with a renewable energy system (i.e., PV panels), an electricity storage system (i.e., battery), as well as an EV.

### 3.1 Building modelling

This study considered a real building cluster located in Ludvika, Dalarna region, Sweden. This building cluster consists of three separate buildings, as shown in Fig. 3. The building cluster (all the three buildings) includes 48 multifamily dwelling units over three floors, and most of the apartments have one or two bedrooms. The total façade surface gross area of the complex is 2146 $m^2$, the total roof surface gross area is 1750 $m^2$. These buildings will be improved by a series of renovation plans including installation of PV, battery storage, direct current (DC) micro grid, and EV charging station. It is assumed the heating is provided by district heating system. So, the PV panels will only need to provide power supply to the domestic electricity demand (e.g. lighting, TVs, dish wash). The schematics of the building cluster and the energy systems are shown in Fig. 1.

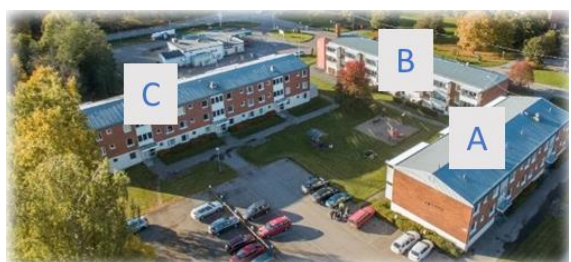

Figure 3 Bird view of the case building cluster located in Ludvika, Sweden

Until now, there are different models developed for modelling the electricity demand in residential buildings. For instance, Palacios-Garcia et al. [38] developed a high-resolution model for calculating the electricity demand of heating and cooling appliances considering variables such as the number of residents, location, type of day (weekday or weekend) and date. In [39], a stochastic model for simulating lighting power consumption profiles in Spain was developed considering the number of household residents and differentiating between weekdays and weekends. In [40], Widén developed a stochastic model for computing the occupancy and electricity load in Sweden. Since the occupancy schedules and lighting usage can vary significantly in different countries due to the culture and location difference, this study chooses Widén's model to calculate the electricity load profiles for the three individual buildings. Meanwhile, in order to achieve acceptable accuracy, the measured data about the annual electricity usage was used to calibrate the model.

### 3.2 Renewable energy system modeling

The power generation from the PV panel $P_{PV}$ ($kW$) is calculated by Eqn. (14) [7, 41],

$$P_{PV} = \tau \times I_{AM} \times I_T \times \eta \times CAP_{PV} \tag{14}$$

where $\tau$ is the transmittance-absorptance product of the PV cover for solar radiation at a normal incidence angle, ranging from 0 to 1; $I_{AM}$ is the combined incidence angle modifier for the PV cover material, ranging from 0 to 1; $I_T$ ($W/m^2$) is the total amount of solar radiation incident on the PV collect surface; $\eta$ is the overall efficiency of the PV array; $CAP_{PV}$ ($m^2$) is the PV surface area.

### 3.3 Electrical battery and EV battery modeling

This study used simplified electrical battery and EV battery models. The electricity stored in the battery is calculated using a simplified model, as expressed by Eqns. (4) and (5). It is estimated from the hourly charging rates [35]. This study considers three EVs. Table 2 summarizes the capacity, maximum charging rates as well as the parking periods of each EV. EV 1, EV 2 and EV 3 are assumed to be charged in Building A, B and C, respectively. To consider the various EV usage, these three EVs are assumed to have different parking periods. EV 1 is assumed to be owned by a resident living in the building, and thus it is parked at night from 18:00 to 07:00 in the next day. EV 2 and EV 3 are assumed to be owned by some working staff in the building estate, and they are parked during daytime (i.e. one from 08:00~16:00 and the other from 09:00~17:00). The EV battery capacity and maximum charging rates are referred from the available EV models in the market in [42].

Table 2 Capacity, charging limits and parking periods of the three different EVs, data obtained from [42]

| ID | Battery capacity (kW·h) | Maximum charging rates (kW) | Parking period |
|---|---|---|---|
| EV 1 | 22 | 4 | 18:00~07:00 (next day) |
| EV 2 | 27 | 5 | 08:00~16:00 |
| EV 3 | 53 | 10 | 09:00~17:00 |

In all the three scenarios, the EVs are required to be fully charged before they leave the charging station. When the EVs arrive in a charging station in the home, a random SOC parameter (between 0 and 1) is assumed to represent the remaining storage in the EV battery.

## 4. Case studies and results analysis

In the case studies, a typical summer week was selected to validate the developed coordinated controls. The weather data of Ludvika was used for modelling the local renewable generations. This section first presents the individual building's electricity demand and renewable generation information. Then, the detailed EV charging and battery charging results obtained from the two scenarios (see Step 4 in Section 2) and the developed control are compared and analyzed. Finally, the overall economic and energy performances are compared.

Table 3 summarizes the input parameters used in the case studies. According to the building dimension, 100 $m^2$, 200 $m^2$ and 300 $m^2$ roof areas are planned for installing PV panels in the three buildings, respectively. It was assumed each building is installed with an electrical battery with capacity of 20 kW·h and a maximum charging/ discharging rates of 6 kW. The price of purchasing electricity from the power grid was set as 0.16 €/(kW·h). Considering the negative impacts on the grid stability and safety, the feed-in-tariff was set as 0.05 €/(kW·h), which is lower than price of electricity purchase [19]. The price of electricity trading in the building cluster was set as 0.1 €/(kW·h). Such price setting will provide incentives for energy sharing within the building cluster, i.e. the building owners can earn more by selling their excessive renewable energy to the building cluster than sell to the power grid, and vice versa.

Table 3 Configuration of the PV and battery system and electricity prices

| Input parameter | Value |
|---|---|
| Area of PV panel in Building A ($m^2$) | 100 |
| Area of PV panel in Building B ($m^2$) | 200 |
| Area of PV panel in Building C ($m^2$) | 300 |
| Battery capacity (kW·h) | 20 |
| Battery maximum charging/discharging rates (kW) | 6 |
| Price of electricity sold to the grid (€/(kW·h)) [19] | 0.05 |
| Price of electricity purchased from the grid (€/(kW·h)) [19] | 0.16 |
| Price of electricity trading in the building cluster (€) | 0.1 |

**4.1 Building electricity demand, renewable generation and electricity mismatch**

Fig. 4 displays the hourly electricity demand, hourly PV generation, and the hourly electricity mismatch of the three buildings in the selected week. Note that the heating needs of the three buildings are assumed to be met by the district heating system. Thus, the electricity demand only includes the domestic electricity loads (i.e. lighting, washing machine, TV, etc.). The trends of PV power production of the three buildings are similar, since the solar irradiation is nearly the same for the three buildings which are located in the same location. As Building C has the largest roof area, more PV panels can be installed on its roof. Thus, it has the largest average PV production.

Power mismatch of each building is calculated as the deviation between the its hourly power demand and hourly renewable generation. A positive value of power mismatch indicates insufficient renewable generation (and thus grid power is needed), while a negative value of power mismatch indicates excessive renewable generation (and thus selling electricity to the grid is needed). The diversity between the power mismatch provides good opportunities for the buildings to collaborative with each other in aspects of energy sharing. For instance, at noon (i.e. 11:00~16:00) in the first day, Building A has insufficient renewable generations (i.e. 7.6 kW·h more demand), while Buildings B and C have excessive renewable generations (i.e. 24.7 kW·h and 55.8 kW·h more supply, respectively). Buildings B and C can share their surplus renewable generation with Building A to avoid grid power imports (for Building A) and power exports to the grid (for Buildings B and C), and thus help improve the overall performance at the building-cluster-level.

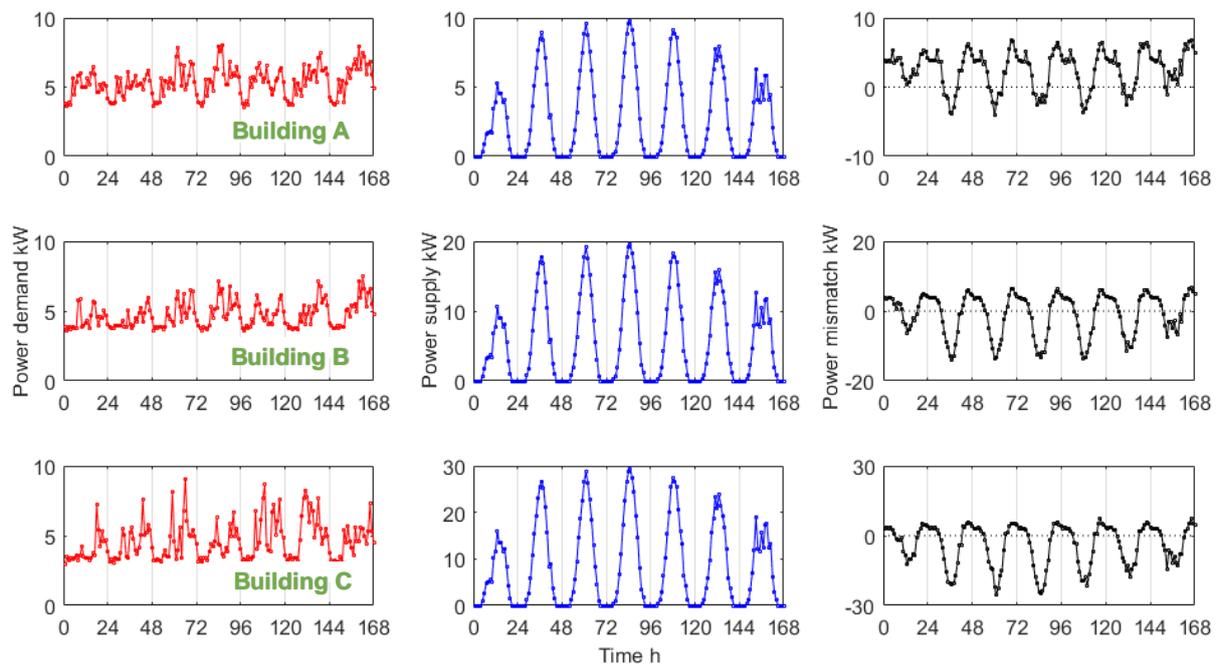

Figure 4 The hourly power demand, renewable generation and power mismatch of three buildings in the selected summer week

### 4.2 Detailed battery controls and energy flows

To have a close look at the charging of EVs and battery storage, as well as the energy flow in the system, the detailed operation in the first day of the selected week is presented and analyzed in this section. Note that the EV charging loads are exactly the same for the three scenarios. The initial SOCs when EVs arrive at the charging stations are the same for three scenarios. The initial SOCs upon arrival for the three EVs are 0.29, 0.61 and 0.62, respectively. All the EVs are required to be fully charged when they depart the charging stations, i.e. SOC equals 1. Fig. 5 presents the State of Charge (SOC) of the three EVs' battery and the aggregated battery in the first day of the selected week. For Scenarios 1 and 2, since the EVs are charged at their maximum charging rates (i.e. 4 kW, 5 kW and 10 kW for the three EVs, respectively) immediately after being plugged into the charging ports, there is a stable increase in the SOCs for all the three EVs in the beginning of parking periods. In the developed control, the EVs are charged flexibly in the parking period. In some timeslots, they are charged at a high rate; while in some timeslots, they are charged at a low rate (or even zero). Despite the different charging patterns, all the EV batteries are fully charged (as specified in the case study, see Section 3.3) before they depart the charging ports in the three scenarios.

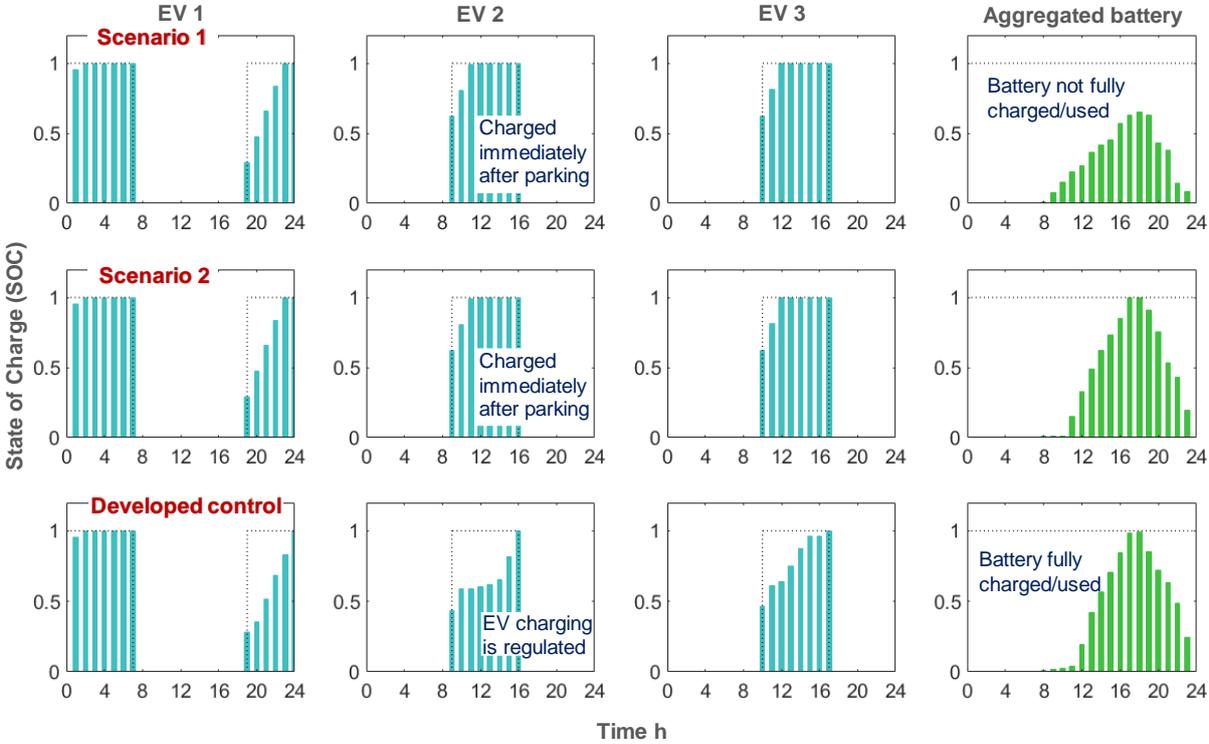

Figure 5 State of Charge (SOC) of the three EVs and the aggregated battery in the first day of the selected week

Regarding the battery storage usage, the aggregated battery has not been fully charged in Scenario 1, while it has been fully charged in Scenario 2 and the developed control. This is because in Scenario 1 the collaboration (i.e. renewable energy sharing) is not allowed among the buildings, while in Scenario 2 and the developed control, collaboration is enabled (see Fig. 6 for detailed energy sharing). The collaboration enables buildings to store their surplus renewables in other building's battery, thereby helping increase the overall battery utilization. Such increased battery utilization can help the building cluster keep more renewable energy onsite instead of exporting to the power grid, and thus contribute to increased renewable energy self-consumption rates.

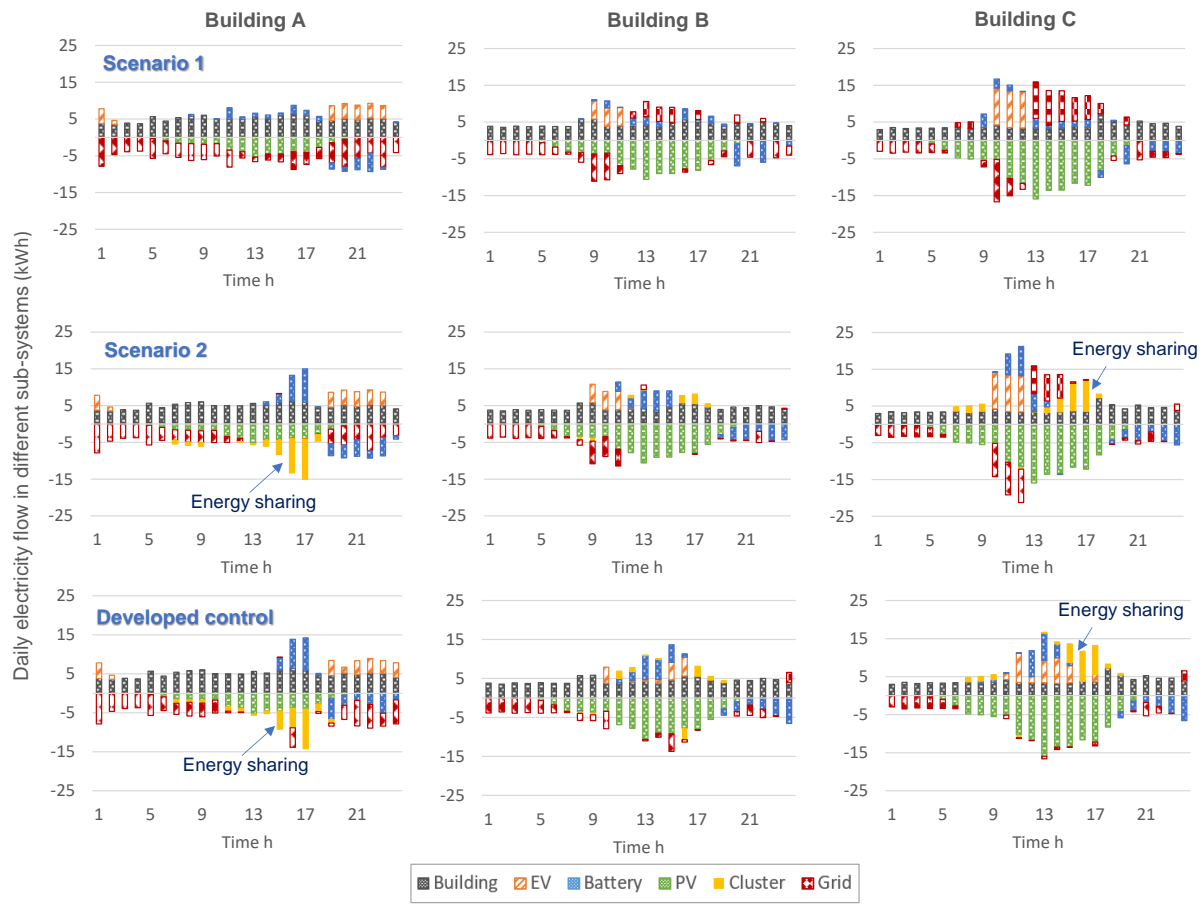

Figure 6 Detailed electricity flow (of building, PV systems, battery and EV) in the individual building in each scenario in the first day of the selected week

Fig. 6 depicts the electricity energy flow of the subsystems (i.e. electrical demands, renewable generation, EV demands and battery charging/discharging) in each building in the first day of the selected week for the three different scenarios. Fig. 6 also shows the individual building's energy exchange with the building cluster (i.e. the other buildings) and with the power grid. A positive value of energy flow indicates energy demand, while a negative value indicates energy supply. For the power grid/building cluster, a positive energy flow indicates buildings export electricity to the power grid/building cluster, while a negative energy flow indicates buildings import electricity from the power grid/building cluster. The PV system produces electricity from 6:00 to 20:00. But for Buildings A and B, the amount of PV power production is less than the electricity demand in the early morning (6:00~8:00). As a result, there is grid power purchase in this period. While Building C has more power production in this period due to a larger PV system installed.

In Scenario 1, the energy sharing is not allowed among the buildings, so there is no energy exchange with the building cluster. Since the EVs are charged immediately after they are plugged into the charging stations, large EV loads can be observed for both Building B and C during 8:00~12:00 (one EV arrives at 9:00, and another EV arrives at 10:00). However, due to a lack of renewable generation in this period, a large amount of grid power is imported from the power grid (i.e. 19.4 kW·h and 27.8 kW·h electricity imports for Buildings B and C, respectively). At noon (i.e. 13:00~15:00), both Buildings B and C have surplus PV power production. However, due to a lack of energy sharing, such surplus PV power is exported to the power grid (i.e. 12.4 kW·h and 27.8 kW·h electricity exports for Buildings B and C, respectively). Meanwhile, Building A purchased 5 kW·h electricity from the power grid. In total, in this, Building A imported 102.7 kW·h electricity, Building B imported 55.6 kW·h electricity and exported 19.7 kW·h electricity, and Building C imported 45.6 kW·h electricity and exported 48.5 kW·h electricity.

In Scenario 2, energy sharing is enabled within the building cluster. At around 14:00, since the battery of Building C has already been fully charged, the surplus renewable energy from Building C is exported to Building A and stored in Building A's battery. Such energy sharing avoids the unnecessary renewable exports to the power grid, as compared with Scenario 1. In total, from 12:00 to 18:00, Building B shared 6.2 kW·h electricity with Building A, and Building C shared 22.3 kW·h electricity with Building A. Similar to Scenario 1, due to a lack of EV control, large EV charging demands can still be observed for both Buildings B and C during 8:00~12:00, leading to large grid power purchase (i.e. 17.8 kW·h and 27 kW·h electricity imports for Buildings B and C, respectively). In total, in this day, Building A imported 71.9 kW·h electricity, Building B imported 41.5 kW·h electricity and exported 1.5 kW·h electricity, and Building C imported 43.2 kW·h electricity and exported 23.4 kW·h electricity.

In the developed control, the energy sharing is enabled among buildings and the EV charging is regulated. Similar to Scenario 2, the renewable energy sharing reduces the renewable energy exports to the power grid from Building C. In terms of EVs charging, due to the shifting of large EV charging loads to periods with more renewable energy generations in Buildings B and C, the developed control effectively reduces the grid power imports in the period 8:00~12:00 compared with Scenario 1 and 2. In total, in this day, Building A imported 74.5 kW·h electricity, Building B imported 40.1 kW·h electricity and exported 2.5 kW·h electricity, and Building C imported 25.1 kW·h electricity and exported 25.7 kW·h electricity.

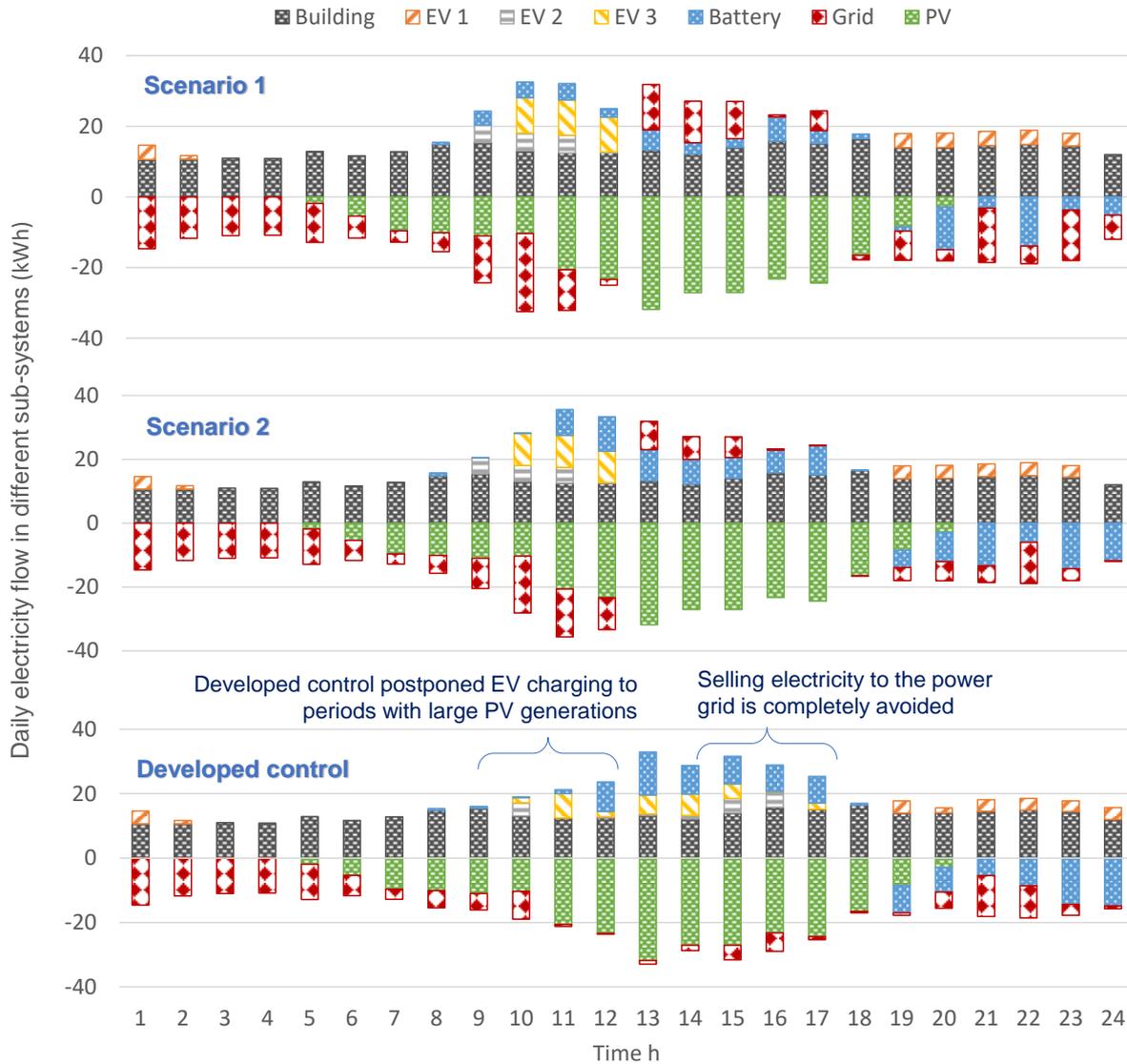

Figure 7 Detailed energy flow (of building, PV systems, battery and three EVs) in the building cluster in each scenario in the first day of the selected week

Fig. 7 depicts the electricity energy flow of the building cluster (i.e. electricity demand), aggregated PV production, power grid, aggregated battery and three EVs in the first day of the selected week for the three different scenarios. The aggregated energy exchanges within the building cluster become zero in the aggregated level, since the amount of purchased electricity from the building cluster compensates with the amount of electricity sold to the building cluster.

In the period 9:00~12:00, for Scenario 1 and Scenario 2, large electricity demand occurs, as EV 2 and EV 3 are charged immediately after being plugged in. Unfortunately, the renewable energy generation is not sufficient in this period to meet the large demands. As a result, a large amount of grid electricity is purchased by the building cluster, i.e. 48.7 kW·h and 52 kW·h for Scenarios 1 and 2, respectively. In Scenario 3 (developed control), as EV 2 and EV 3 can be flexibly charged in any timeslot during the parking period, the controllers set relatively small

EV charging rates in this period. Consequently, the amount of grid power purchase is significantly reduced in the developed control, i.e. 14.6 kW·h. In the period 14:00~17:00, for Scenario 1, since there is no collaboration among buildings, only a small part of the surplus renewable energy is kept onsite, while a large part of the surplus renewables (i.e. 28.5 kW·h) is exported to the power grid at a low price. In Scenario 2, contributed by the energy sharing within building cluster, more renewable energy can be stored in the battery. After the batteries in the building cluster all being fully charged, only a small amount of surplus renewable energy (i.e. 14.1 kW·h, which is only half of the amount of exported electricity in Scenario 1) is still exported to the power grid. Scenario 2 has better performance compared with Scenario 1. Since the batteries of EV 2 and EV 3 have already been fully charged in the period 9:00~12:00, there is no energy flow for them in the period 14:00~17:00. In the developed control, considering the large renewable energy production in this period, the controller shifts the charging load of EV 2 and EV 3 to this period. Part of the surplus renewable generation is stored in the building battery and part of the surplus renewables is used to supply the EV load. As a result, exporting renewable energy to the power grid is completely avoided. This can effectively improve the renewable energy self-consumption rate of the building cluster.

To sum up, in Scenario 1, the building cluster exported 41.3 kW·h electricity to the grid and imported 177.0 kW·h electricity from the grid. In Scenario 2, the building cluster exported 23.0 kW·h electricity to the grid and imported 159 kW·h electricity from the grid. Scenario 2 performs better than Scenario 1 (i.e. with reduced energy imports/exports) as energy sharing enables the building cluster to keep more renewable energy on-site. While using the developed control, the building cluster exported 0 kW·h electricity to the grid and imported 135.6 kW·h electricity from the grid. Scenario 3 performs even better than Scenario 2, as the controller shifts EV charging loads to periods with large renewable production and thus help keep more renewable energy used onsite in cases when the batteries in the buildings have been fully charged (see the aggregated battery SOC in Fig. 5).

### 4.3 Overall economic and energy performance comparison

This section compares the overall economic and energy performance of different controls. Table 4 summarizes the building-cluster-level daily electricity costs and renewable energy self-consumption rates. in different scenarios. Fig. 8(a) compares the daily renewable energy self-consumption rates of the three scenarios in the selected week. The relative performances improvements of Scenario 2 and the developed control compared with Scenario 1 are also depicted. Compared with Scenario 1, Scenario 2 improved the renewable energy self-consumption by 6%~9%. This is because the collaboration enables buildings to share their surplus renewable energy with other buildings with insufficient supply and thus help reduce the electricity exports to the power grid (i.e. keep more renewable energy onsite). Compared with Scenario 2, the developed control further improves the renewable self-consumption rates by 3% to 11% (see Table 4). This is because the developed control makes use of the flexible charging ability of EVs. By shifting the EV charging load to periods with large renewable generation periods, more renewable energy can be used onsite, especially when the electrical battery storages are fully charged. In Day 1, the daily self-consumption rates are relatively higher than the other six days. This is because the amount of PV power production is relatively lower than other days (see Fig. 4 the PV power production profiles), and thus most of the PV power will be used to meet the electricity demand on-site.

Table 4 Building-cluster-level daily economic and energy performance in different scenarios

|  | Day | 1 | 2 | 3 | 4 | 5 | 6 | 7 |
|---|---|---|---|---|---|---|---|---|
| Daily self-consumption | Scenario 1 | 83.7% | 60.2% | 60.9% | 59.9% | 65.1% | 62.7% | 75.2% |
|  | Scenario 2 | 91.0% | 64.2% | 65.5% | 63.4% | 70.4% | 67.3% | 80.3% |
|  | Proposed control | 100.0% | 66.2% | 68.3% | 65.5% | 74.4% | 73.2% | 86.8% |
| Relative improvements | Scenario 2 vs 1 | 9% | 7% | 8% | 6% | 8% | 7% | 7% |
|  | Developed vs 1 | 19% | 10% | 12% | 9% | 14% | 17% | 15% |
| Daily electricity costs (€) | Scenario 1 | 29.2 | 13.7 | 14.5 | 12.2 | 15.1 | 15.4 | 20.4 |
|  | Scenario 2 | 24.2 | 9.9 | 12.2 | 9.8 | 13.4 | 11.2 | 15.4 |
|  | Proposed control | 21.7 | 9.0 | 10.8 | 8.6 | 12.4 | 8.5 | 13.0 |
| Relative improvements | Scenario 2 vs 1 | 17% | 27% | 16% | 20% | 11% | 28% | 25% |
|  | Developed vs 1 | 26% | 34% | 25% | 29% | 18% | 45% | 36% |

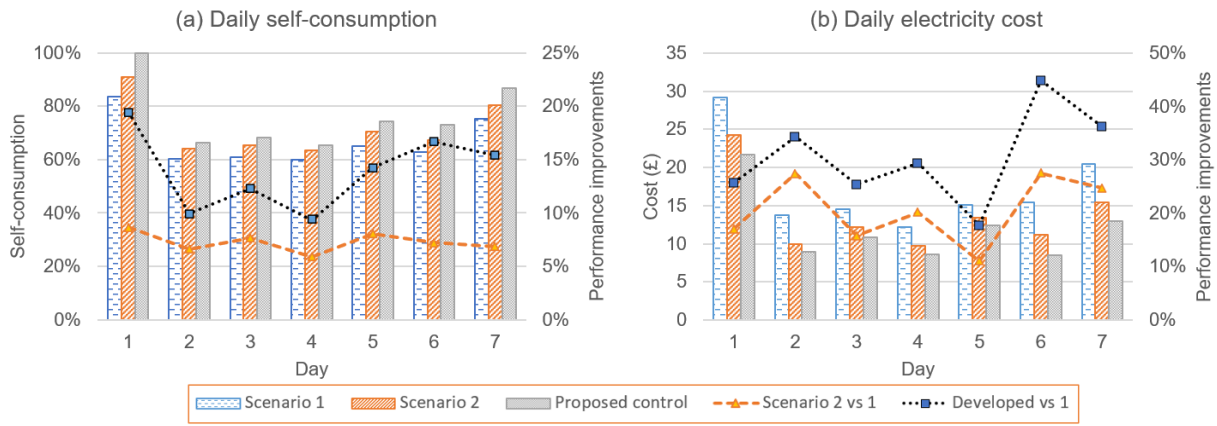

Figure 8 Comparison of the daily renewable energy self-consumption rates and daily electricity costs of the three scenarios

Fig. 8(b) compares the daily electricity costs of the three scenarios in the selected week. Due to increased renewable energy self-consumption rates and thus less grid power purchase, Scenario 2 achieves 11%~28% cost saving compared with Scenario 1, and the developed control achieves 7%~17% more cost saving compared with Scenario 2 (see Table 4). The relative improvements in economic performance is much larger than the relative improvements in daily self-consumption rates. This is because the building cluster purchase electricity from the power grid at a high price (i.e. 0.16 €/(kW·h)) but sell electricity at a much lower price (i.e. 0.05 €/(kW·h)). When the building cluster exports more renewables to the power grid (i.e. in Scenario 1), they will need to buy more electricity from the grid at a high price, as the aggregated daily electricity demand is fixed.

## 5. Conclusion

This study has proposed a coordinated control of building clusters for improving the cluster-level performance, with both energy sharing and EV charging considered. The developed coordinated control first uses a 'representative' building to represent the whole building cluster and optimizes its energy storage operation as well as the EV charging using genetic algorithm. The optimized performance of the building cluster is considered to be the optimal one that maximizes the energy sharing within the building cluster by coordinating individual building's operation. Then, non-linear programming is used to coordinate the operation of each individual building. For validation, the developed control has been tested using the energy demand and supply data on a real buildings cluster (with three EVs considered) in Ludvika, Sweden, and its detailed energy performance (i.e. renewable self-consumption rate) and economic performance (i.e. electricity cost) have been compared with two scenarios (i.e. one does not enable energy sharing and one allows full energy sharing, both do not have EV charging controls). Case studies have shown that the developed coordinated control can effectively improve the renewable self-consumption rates and meanwhile reduce the electricity bills of the building cluster, by taking advantage of energy sharing, storage capability of electricity batteries, flexible demand shifting ability of EVs. The major findings are summarized as follows:

- The developed coordinated control provides a mechanism to coordinate each single building's operation and EV charging demands for improved building cluster performances.
- In aspect of renewable utilization, the coordinated control improved the daily self-consumption rates by as much as 19% compared with Scenario 1 (no EV control and no energy sharing) and as much as 11% compared with Scenario 2 (no EV control but with energy sharing). This is because the developed control shifts the EV charging load to periods with large renewable generation periods, and thus more renewable energy are used onsite, especially when the electrical battery storages are fully charged.
- In aspect of economic costs, the coordinated control reduced the daily electricity costs by as much as 36% compared with Scenario 1 (no EV control and no energy sharing) and as much as 17% compared with Scenario 2 (no EV control but with energy sharing). This is because the developed control reduces the amount of high-price grid electricity imports.

This paper concentrates on the development of control concept for the coupled PV-battery storage-EV systems in the case building cluster. So far, the renovation in this demo case is still under progress and the only pre-monitoring data before renovation has been collected. In the future with both PV and EV integrated, post-monitoring data will be collected and the experimental data from the demo site will be used to validate the simulation. In this study, the detailed driving patterns of EVs are not considered, and the SOC when they arrive the charging ports are determined by some random values. The mobility of humans is highly regular, and study shows that there is 93% potential predictability in user mobility [43]. Future work will take account of the predictive EV driving patterns in the optimization to achieve better performances. Meanwhile, the uncertainty in demand and renewable prediction is not considered in this study. Future work will try to develop more robust controls.

## Acknowledgments

The authors are thankful for the financial support from EU Horizon 2020 EnergyMatching project (Grant no. 768766), the UBMEM project from Swedish Energy Agency (Grant no. 46068) and the European Regional Development Fund and Region Dalarna through the project Energiinnovation.